\documentclass[a4paper,twocolumn]{article}

\usepackage{graphicx} 
\usepackage[english]{babel}
\usepackage{times} 
\usepackage{amsmath}
\usepackage[hyphens]{url}
\usepackage{hyperref} 
\usepackage{booktabs} 
\usepackage{eurosym}
\usepackage[numbers]{natbib}
\usepackage{flushend}
\usepackage{csquotes}

\usepackage{subfig}
\usepackage{pgfplots}
\pgfplotsset{compat=1.14}

\title{On optimising cost and value in compute systems for radio astronomy\thanks{Preprint accepted for publication in Astronomy and Computing}} 

\begin{document}

\author{P. Chris Broekema\thanks{corresponding author, email: \href{mailto:broekema@astron.nl}{broekema@astron.nl}} \\
   ASTRON \and
   Verity Allan\\
   Universit of Cambridge \and
   Rob V. van Nieuwpoort\\
   Netherlands eScience Center and \\
   University of Amsterdam \and   
   Henri E. Bal\\
   Vrije Universiteit Amsterdam}



\maketitle

\begin{abstract}
Large-scale science instruments, such as the distributed radio telescope LOFAR, show that we are in an era of data-intensive scientific discovery.
Such instruments rely critically on significant computing resources, both hardware and software, to do science.
Considering limited science budgets, and the small fraction of these that can be dedicated to compute hardware and software, there is a strong and obvious desire for low-cost computing.
However, optimising for cost is only part of the equation; the value potential over the lifetime of the solution should also be taken into account.
Using a tangible example, compute hardware, we introduce a conceptual model to approximate the lifetime relative science value of such a system.
While the introduced model is not intended to result in a numeric value for merit, it does enumerate some components that define this metric.
The intent of this paper is to show how compute system related design and procurement decisions in data-intensive science projects should be weighed and valued.
By using both total cost and science value as a driver, the science output per invested Euro is maximised.
With a number of case studies, focused on computing applications in radio astronomy past, present and future, we show that the hardware-based analysis can be, and has been, applied more broadly.

\end{abstract}


\section{Introduction}
\label{sec:intro}
Modern large-scale science instruments critically rely on specialised data-intensive computer technologies, to turn instrument data into useful science results.
Considering limited science budgets, of which only a small fraction can be dedicated to computing, there is a strong desire to use these expensive systems in an optimal way.
The design of such an optimised system is heavily influenced by experience from previous installations.
For instance, the design priorities of the GPU-based correlator and beamformer system for the LOFAR radio telescope, in particular its focus on an I/O optimised design, borrowed heavily from previous experiences with Blue Gene based systems~\cite{Broekema:18a}.

In this paper we discuss both the cost and value of computing technologies, and how to optimise the combination of these two for maximum science impact.
Since these are difficult to measure for the complex combination of hardware, middleware and software that are generally required, we focus our detailed analysis on hardware.
We enumerate some of the factors that impact the total cost of a system. 
However, we propose that total cost over the lifetime of a system is only part of the equation: the computational and scientific performance of different solutions may radically differ for the applications in question, depending on system and application characteristics. 
A more valuable metric would look at the useful output of a system per invested Euro.
For example, the Distributed ASCI Supercomputer (DAS)~\cite{Bal:2016} consortium tracks the effectiveness of its distributed cluster infrastructure via the number of awarded PhDs per cluster generation, as shown in Table \ref{tab:das-phds}\footnote{source: \url{https://www.cs.vu.nl/das4/phd.shtm}, \\\url{https://www.cs.vu.nl/das5/phd.shtml} and historical data}.
Considering the nearly constant budget for these systems, between 1.2 and 1.5 M\euro, discounting inflation, the cost per supported PhD has dropped considerably over the lifetime of the DAS consortium.
Alternatively, we can argue that the relative science value per invested Euro has dramatically increased.

\begin{table}[hbt]
  \centering
  \resizebox{\columnwidth}{!} {
    \begin{tabular}{ccccl}
            & Year & PhDs & \euro / PhD & Research agenda \\
      \cmidrule(r){2-5}
      DAS-1 & 1997 & 7    & \euro ~214.285 &  Wide-area computing \\
      DAS-2 & 2002 & 22   & \euro ~~68.181 &  Grid computing  \\
      DAS-3 & 2006 & 36   & \euro ~~41.666 &  Optical grids   \\
      DAS-4 & 2010 & 33   & \euro ~~45.454 &  Clouds, diversity, green IT  \\
      DAS-5 & 2015 & 40   & \euro ~~37.500 &  Harnessing diversity \& complexity \\
      \\
    \end{tabular}
  }
  \caption{Awarded PhDs per Distributed ASCI Supercomputer generation}
  \label{tab:das-phds}
\end{table}

In this paper we study a number of cases in radio astronomy, a computationally- and data-intensive science that has been using high-performance computing technologies since the very early days of computing to achieve scientific results.
We show how the methodology proposed in this paper has informally been used in the past.
The main contributions in this paper are:
\begin{itemize}
\item the introduction of the concepts \textit{relative science value} and \textit{total value of ownership}, including two potential ways to estimate total value of ownership over the lifetime of a system,
\item the introduction of a way to reason about compute system technology beyond just cost,
\item a number of case studies that show practical trade-offs between cost and value in radio astronomy.
\end{itemize}

Although we present a number of equations in this paper, it is not our intention that these are used to generate a numeric merit value for a particular system or technology.
Rather, they are intended to illustrate which components contribute to the cost and merit of a system and as a starting point for a more detailed discussion on the relative value of various compute systems.
With these components, and some examples of cost and value past and present in this paper in mind, system designers and architects have the tools needed to better balance their designs, and evaluate their design choices within this framework.

The intent of this paper is to show how compute system related design and procurement decisions in data-intensive science projects should be weighed and valued.
By using both total cost and science value as a driver, the science output per invested Euro is maximised.
While the general concepts discussed in this paper are known in systems engineering, we hope to introduce them to a broader audience of scientific decision makers, principal investigators, and system architects and designers.


\section{Compute systems for large-scale science}





The study of Physics, in particular Astrophysics, has relied on state-of-the-art computer science and high-performance computing.
Modern aperture synthesis radio astronomy in particular was made possible by the development of the Fast Fourier Transform (FFT)~\cite{Cooley:1965}\footnote{Ryle in his Nobel lecture credits Dr. David Wheeler with the invention of the FFT in 1959~\cite{Ryle:1974}} and computers fast and cheap enough to use them at scale.
For example, the One-Mile Telescope, built at the Mullard Radio Astronomy Observatory, Cambridge, in 1964, relied on the computing advances of the EDSAC II and TITAN computers, as is illustrated in our Case Studies in Section \ref{sec:Titan}.
This telescope, and others, like the Half-Mile Telescope at Cambridge and the Westerbork Synthesis Radio Telescope in the Netherlands, depended on the abundant and increasingly cheap computation available to develop the new scientific technique of aperture synthesis, which unlocked new science and ultimately won a Nobel Prize. 

More recently, the range of applications that benefit from large-scale computing has increased dramatically with the rise of Data Science, and the ease with which high-performance (if not world-leading) compute infrastructures have become available via Cloud Computing.
This paper is thus presented at a timely moment, to provide decision makers, principal investigators and designers of new compute systems and applications with a framework to help evaluate and guide their design choices. 

\section{On relative science value}
\label{sec:merit}
In the previous section we have argued that modern data-intensive science relies heavily on computing.
Given the high cost of such resources, there is an obvious desire to maximise their usefulness, or minimise their cost.
We introduce a system's \textit{Relative science value}, defined as its value per invested Euro over its lifetime, as a measure for the merit of a system over its lifetime.
The definition of \textit{value} will be discussed in Section \ref{sec:tvo}.

The computational systems supporting modern data-intensive science are often a complex collection of hardware, middleware and software.
Quantifying the cost and relative value of such a complex integrated system is nearly impossible.
To start our exploration we will focus on one of the more tangible components: hardware.

By first exploring ways to quantify hardware cost and value, we reduce the complexity of the system under investigation without impacting the value of the analysis.
In section \ref{sec:case_studies} we show that the methodical hardware-based analysis can be applied more broadly, as similar considerations can be used to evaluate other system costs, such as software development, maintenance and power consumption.

The relative usefulness of a hardware system, its relative science value ($M_S$), depends on its total aggregate value accrued over time (total value of ownership, $TVO$) and aggregate cost over the lifetime of the system (total cost of ownership, $TCO$):
\begin{equation}\label{eq:main}
M_S=\frac{TVO}{TCO}
\end{equation}

Total Cost of Ownership is a well known concept, both as a tool to inform purchasing decisions in general~\cite{ellram:1995}, and in computer science.
In this paper we give our own definition of the Total Cost of Ownership of a system.
We introduce the generic concept of Total Value of Ownership in this paper.

From Equation \ref{eq:main} it is obvious that there are two ways to maximise the relative science merit of a compute system: reduce Total Cost of Ownership, or increase Total Value of Ownership of a system.
In practice, a carefully considered combination of the two is likely to produce the optimal result.
Obviously, total cumulative value $TVO$ is not easy to quantify, and we note that the time over which value is accumulated may extend well beyond the lifetime of the system.
In the next sections we will explore the components that make up $TVO$ and $TCO$.





\section{Total Value of Ownership}
\label{sec:tvo}
Whereas the concept of Total Cost of Ownership is well known and established, the same can not be said for its value counterpart; we shall therefore introduce this first.
In economic terms, we are interested in the return on investment, which we'll refer to as Total Value of Ownership (TVO) in this paper, to contrast to Total Cost of Ownership.
While this is an essential question to ask during the definition phase of a project, the answer is seldom easy to quantify.
The success of science projects is generally measured in the importance of its scientific results, often expressed in the number of published peer-reviewed papers produced.
However, from a system design perspective, it is attractive to use a more easily measured metric, such as compute power, throughput or storage capacity, to describe the value of a system.
While such metrics are convenient and may be useful in their own right, we argue that these do not necessarily provide an accurate reflection of how the system will be used.
Furthermore, these do not necessarily take computational efficiency, scientific impact, or average required capacity per accepted paper into account.
In this section we propose two measures for a system's TVO that are designed to more accurately reflect the actual scientific usefulness of a system: total lifetime computational value ($V_c$), and total lifetime scientific value ($V_s$).
While we provide equations, these are not designed to be used to model TVO; but rather to capture the relationship between some of the various elements that define system value.

\begin{figure}
  \includegraphics[width=\columnwidth]{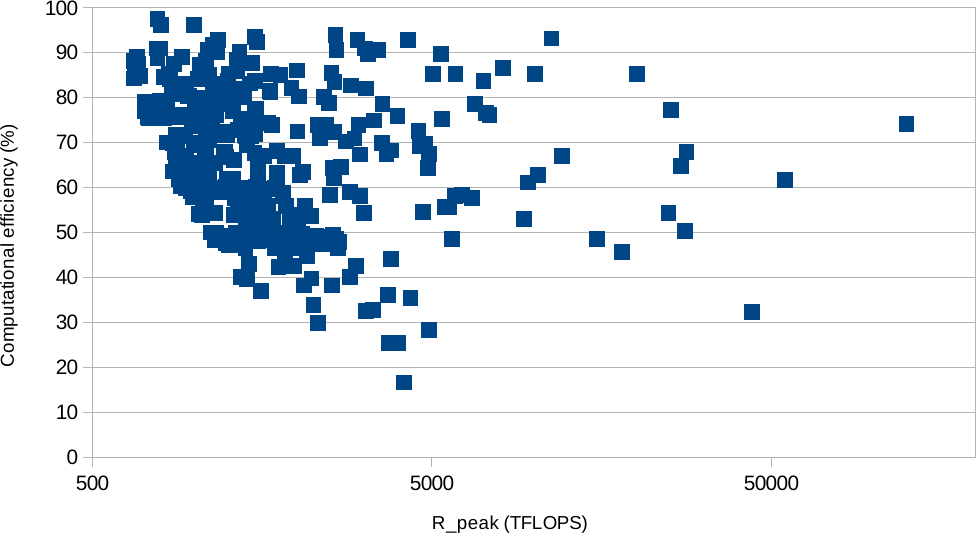}
  \caption{HPL computational efficiency in the Top500 (November 2017)}
  \label{fig:top500efficiencies}
\end{figure}

Total performance, computational or otherwise, of a system can be a useful measure for the value of a (hardware) system.
However, even this can be difficult to quantify beforehand.
Whereas peak computational performance is relatively easy to determine, often only a small fraction of this can be achieved in practice.
The same can be said for other metrics like peak network and storage performance.
The fraction of the computational resources that can effectively be used by an application is determined by its computational efficiency.
A discussion on the factors that impact computational efficiency is beyond the scope of this paper, but we note that these factors should be foremost in the mind of a hardware system architect.
To illustrate this point, we look at the yearly Top500 list of the fastest supercomputer in the world for the HPL benchmark~\footnote{\url{www.top500.org}}.
Computational efficiencies of these systems, shown in Figure \ref{fig:top500efficiencies}, range from $15.6\%$ to $97.6\%$, which shows that the impact of unexpectedly low computational efficiency may be catastrophic.

By taking into account the target applications for a specific system, we introduce an estimate for its total lifetime computational value ($V_c$, in FLOP), as shown in Equation \ref{eq:computational_value}.
\begin{equation}\label{eq:computational_value}
  V_{c} = T_l ~ A_{o} ~ \sum_{p=0}^{P}{ \Big( f_p ~ R_{max,p} \Big) } \text{, with} \sum_{p=0}^{P}{f_p \leq 1}
\end{equation}
Here, we take the total lifetime of the system, $T_l$, and its availability as a fraction of total lifetime, operational availability ($A_o$), to get the effective time the system is usefully available over its lifetime.
For each application $p$, its maximum achieved performance on the target system ($R_{max,p}$), and the fraction of operational time it is expected to be run ($f_p$) are taken to get a value for the average maximum achieved performance over all applications to be run on the system.
Combined, these two components make up the system's total lifetime computational value.

Equation \ref{eq:computational_value} thus takes computational efficiency into account over all target applications, and considers both system lifetime and operational availability.
Similar analyses could be done for other performance metrics, such as network bandwidth and storage system performance.
We do not measure average performance of applications, rather we determine the total effective performance over the lifetime of the system.
However, the eventual goal of a compute system is not the delivery of capacity per se, but rather to facilitate science.
A discussion on appropriate metrics for scientific output is well out of scope for this paper, for the purpose of the discussion in this paper we use \textit{scientific publication} as a placeholder.
This is most easily measured in peer-reviewed journal or conference publications; however, one may also consider monographs and PhD theses, or even awards (see section \ref{sec:Titan}).

To illustrate these points, we introduce a system's \textit{Total lifetime scientific value}, which in our example is based on its previously introduced total lifetime computational value.
Since not all science requires the same amount of resources, processing power or other, per scientific publication, we add the \textit{average computational resource required per scientific publication}.
An appropriate \textit{impact factor}\footnote{We are aware that constructing a useful impact factor has many pitfalls. See \cite{Lariviere:18} for a discussion about journal impact factors as an example.}, which is not necessarily the same as a journal impact factor, may be added to differentiate potential Nobel prize winning research from more generic projects.
Notably, this impact factor may be highly time sensitive, in the sense that ground-breaking projects generally have very high impact factors (see section \ref{sec:lofar} for an example).
We note that these two factors may be subjective, highly sensitive, and may have significant political implications.

\begin{equation}\label{eq:scientific_merit}
  V_s = T_l ~ A_o ~  \sum_{p=0}^{P}{ \Big( f_{p} ~ \frac{R_{max,p}}{C_{cpp,p}} ~ I_p \Big) } \text{, with} \sum_{p=0}^{P}{f_{p} \leq 1}
\end{equation}

Total lifetime scientific value $V_s$ is defined in Equation \ref{eq:scientific_merit} by the maximum achieved performance of the application associated with science case $p$ on the system under investigation $R_{max, p}$, divided by the average amount of resources required per scientific publication for that science case $C_{cpp,p}$.
This results in the number of scientific publications per unit of time for that science case and system.
Multiplied by some impact factor per science case, $I_p$, and summed over all science cases targeted by the system $P$ and normalised using the fraction of time each application is expected to consume ($f_p$), gives us a measure for scientific impact per measure of time for that system.
Multiplying that by the total lifetime of the system $T_l$ and the fraction of that time the system is actually available (operational availability $A_o$) gives us the total scientific value of a system, in a unitless \textit{scientific impact}.
For convenience we use computational resources, in floating point operations (FLOP), as a measure for resources required per scientific publication in this model, but other metrics (such as bandwidth, storage capacity, etc), or a combination of such metrics, may be used instead.

The two value measures introduced in this section are by no means the only ones that can be defined.
They are intended to start the discussion and offer an initial indication of the processes and thinking involved.
Characterising the performance of a compute system in a single number is notoriously difficult, which has been studied in some detail.
Previous work suggested the use of harmonic means of runtime of a number representative benchmarks to express the useful performance of a computer~\cite{smith:1988}, which expresses performance in terms of the total runtime of a set of benchmarks.
While benchmarks certainly have their place, and runtime is an appropriate measure for performance of a system, this is not necessarily equivalent to value.
However, we note that the $V_c$ is equivalent to the weighted  harmonic mean suggested by Smith et al. multiplied by $T_l ~ A_{o}$.
Essentially, instead of computing average performance, we focus on aggregate performance over the effective lifetime of the system, taking system lifetime and operational availability into account.






\section{Total Cost of Ownership}
\label{sec:tco}
Having looked at various ways to define the value potential of a system, we now turn to more familiar ground: cost.
The aggregate cost of a system over its lifetime is usually referred to as its Total Cost of Ownership.
While the definition of TCO is relatively easy to give, calculating it a priori may not be as simple, in particular in large-scale science installations.
The lifetime of a particular system may be unpredictable, and the often non-conventional use of such systems may lead to unexpectedly large operational costs.
Furthermore, complex and highly integrated systems make for difficult deployment and integration, which is hard to plan and budget for.
Having said that, TCO can be defined as a combination of capital investment ($C_{cap}$), engineering cost ($C_{eng}$, often called non-recurring expense, or NRE), installation, deployment and integration cost ($C_{int}$), development cost ($C_{dev}$), recurring operational cost ($C_{ops}$) over the lifetime of the system ($T_{l}$) and miscellaneous costs not covered elsewhere $C_{misc}$, as shown in Equation \ref{eq:tco}.
\begin{equation}\label{eq:tco}
TCO = C_{cap} + C_{eng} + C_{int} + C_{dev} + \sum_{t=0}^{T_{l}}{C_{ops}} + C_{misc}
\end{equation}

The one time investment to acquire a system is referred to as its capital cost, $C_{cap}$.
This includes all readily available hardware required to install and commission the system.
Capital cost is usually either capped, or relatively easy to estimate.
We note, however, that even capital cost becomes highly uncertain when predicted several years in advance, due to fast moving markets and uncertain performance characteristics and pricing of newly developed components.
Models often resort to extrapolation from existing systems using some form of Moore's law scaling to estimate future cost and performance (see for instance the SDP costing for the SKA telescope~\cite{alexander:2013}).
While this has historically been somewhat accurate, the demise of Dennard scaling~\cite{dennard:1974} around 2005 has made modelling much more complicated.
This uncertainty is exacerbated by an erratic market that is increasingly dominated by single players without significant competition.

When a system requires engineering investment in order to be usefully employed, this is engineering cost, $C_{eng}$.
This may involve custom cooling solutions, or other non-standard equipment specific to the system (see for an example the LOFAR GPU-based correlator and beamformer~\cite{Broekema:18a}).
Costs associated with certification of a custom solution may also be considered engineering cost.
General purpose systems generally have no or very little engineering cost, but in more specialised systems this may be a significant cost driver.

Any investment needed to integrate and commission the system into an existing infrastructure is captured in integration and commissioning cost, $C_{int}$.
Note that in software systems, especially if the source code of this software is available, integration, commissioning and development may be closely related.

It is unlikely that the application software of a science instrument or experiment remains static over the lifetime of the instrument.
Part of the software evolution will be to add additional functionality or implement advances in algorithmic or scientific understanding of the problem.
Another part of this development will be to adapt existing code to run (efficiently) on a newly installed platform.
The cost of this particular development effort is the development costs of that system ($C_{dev}$).
Such costs may be small (e.g., porting code to a newer system with the same or a similar architecture), or very large, for example, porting functionality from a CPU cluster to a GPU-based system, as was done for LOFAR correlator~\cite{Broekema:18a}.
These costs may be difficult to predict during the design phase of a long-lived instrument, which, in the LOFAR case, was a decade earlier.
It is likely that not all development effort is expended before the system is deployed, and $C_{dev}$ may extend significantly into the lifetime of the system.
Furthermore, added development effort may have a significant impact on computational efficiency, with a corresponding effect on Total Value of Ownership.
There is a direct coupling between the development costs, and thus Total Cost of Ownership, and Total Value of Ownership.
Conversely, if a system performs well enough, there is no need to expend more development effort to improve performance, unless this opens opportunities for, for instance, additional science cases.

Whereas all previously mentioned costs, with the exception of Development costs, are expended before the system becomes operational, Operational cost ($C_{ops}$) is a recurring line-item during the lifetime of the system.
This includes costs associated with energy consumed, infrastructure cost (i.e. rack space, network connectivity, both physical links and bandwidth, heat dissipation, etc), maintenance and system administration.
We have simplified our model by using a single operational cost component; reality is often more complex, especially in a hosted environment where the components mentioned above are provided by different entities or organisations.
While we have opted to keep operational cost in our model flat over the lifetime of the system, this is again a simplification. 
Operational cost in the initial phase of the system may be higher both due to early failure of hardware and staff unfamiliarity and training.
Near the end of the operational lifespan of the system, often after four or five years in general purpose computing, an increase in hardware failures may be observed, which may increase operational cost, depending on the chosen service model.
Furthermore, operational cost may depend on inherently volatile pricing of, for instance, electricity.
Energy costs are often estimated using the previously mentioned extrapolation using Moore's law scaling, while staffing levels and costs may be based on industry standard fractions of FTE per rack or PetaByte~\cite{graser:2015}.

Finally, staff costs not included in the components above, such as those required to secure funding, acquire the system (e.g. writing tender documentation and evaluating responses) and to decommission the system after its useful lifetime, as well project management and support other than system administration, are included in miscellaneous cost ($C_{misc}$).

The remainder of this paper takes the concepts introduced, and shows, using artificial and real-world case studies taken from radio astronomy past and present, the value of this structured approach to compute system design.

\section{A synthetic instructive example}
\label{sec:experiment}

In the previous sections we identified a metric that we can optimise for: total relative science value as defined in Equation~\ref{eq:main}, $M_s$, but its definition is (deliberately) ambiguous.
While it is not our intention to advocate numeric values for the total relative science values for eScience technologies, we can use the equations introduced above to identify ways to optimise their usefulness.

In this section we illustrate the value of the proposed methodology using a thought experiment.
We have constructed an example that is obviously manipulated to show the desired results.
However, using this example we show that, depending on the value measure selected, any of the offered solutions can be judged superior to the others.



Table \ref{tab:offers} describes hypothetical responses to a hypothetical request for tender for the replacement of key computer hardware.
A set of ten key applications was identified that cover the lifetime of this system, and performance of each system for these applications was measured, as shown in Table \ref{tab:performance-of-offers}.

\begin{table}[hbt]
  \resizebox{\columnwidth}{!} {
    \begin{tabular}{lrrrrr}
      & Cheap & Inefficient & Ops & Custom & Specialized \\ 
      \cmidrule(r){2-6}
      
      $C_{cap}$ (\euro) & 250.000 & 350.000 & 350.000 & 300.000 & 400.000 \\
      $C_{eng}$ (\euro) & - & - & 25.000 & - & 25.000\\
      $C_{int}$ (\euro) & 25.000 & - & 25.000 & -  & 25.000\\
      $C_{dev}$ (\euro) & 750.000 & 600.000 & 1.250.000 & 1.250.000 & 1.000.000\\
      $C_{ops}/yr$ (\euro) & 50.000 & 25.000 & 75.000 & 25.000 & 25.000\\
      $C_{misc}$ (\euro) & 25.000 & 25.000 & 25.000 & 25.000 & 25.000 \\
      $T_l$ (yr) & 5 & 5 & 5 & 5 & 5 \\
      $A_o$ & 0,9 & 0,95 & 0,85 & 0,95 & 0,95 \\
      \cmidrule(r){2-6}
      TCO (\euro)& 1.300.000 & 1.100.000 & 2.050.000 & 1.700.000 & 1.600.000 \\
      \cmidrule(r){2-6}
    \end{tabular}
  }
  \caption{The offered solutions, with detailed cost, lifetime and availability information}
  \label{tab:offers}
\end{table}

\begin{table}

  \resizebox{\columnwidth}{!} {
    \begin{tabular}{lccc|ccccc}
      & $f_p$ & $ C_{cpp}$ & $I_p$ & Cheap & Inefficient & Ops & Custom & Specialised \\
      \cmidrule(r){2-9}
      A & 0,04 & $1\cdot10^4$ & 5   & $2\cdot10^8$ & $1\cdot10^8$ & $5\cdot10^8$ & $4\cdot10^8$ & $2,5\cdot10^8$ \\
      B & 0,08 & $1\cdot10^4$ & 5   & $2\cdot10^8$ & $1\cdot10^8$ & $5\cdot10^8$ & $4\cdot10^8$ & $2,5\cdot10^8$\\
      C & 0,02 & $1\cdot10^4$ & 5   & $2\cdot10^8$ & $1\cdot10^8$ & $5\cdot10^8$ & $4\cdot10^8$ & $2,5\cdot10^8$\\
      D & 0,02 & $1\cdot10^4$ & 5   & $2\cdot10^8$ & $1\cdot10^8$ & $5\cdot10^8$ & $4\cdot10^8$ & $2,5\cdot10^8$\\
      E & 0,40 & $1\cdot10^4$ & 5   & $2\cdot10^8$ & $1\cdot10^8$ & $5\cdot10^8$ & $4\cdot10^8$ & $2,5\cdot10^8$\\
      F & 0,11 & $1\cdot10^4$ & 5   & $2\cdot10^8$ & $1\cdot10^8$ & $5\cdot10^8$ & $4\cdot10^8$ & $2,5\cdot10^8$\\
      G & 0,07 & $1\cdot10^4$ & 5   & $2\cdot10^8$ & $1\cdot10^8$ & $5\cdot10^8$ & $4\cdot10^8$ & $2,5\cdot10^8$\\ 
      H & 0,08 & $1\cdot10^4$ & 5   & $2\cdot10^8$ & $1\cdot10^8$ & $5\cdot10^8$ & $4\cdot10^8$ & $2,5\cdot10^8$\\
      I & 0,02 & $1\cdot10^4$ & 100 & $2\cdot10^8$ & $1\cdot10^8$ & $5\cdot10^8$ & $4\cdot10^8$ & $10\cdot10^8$\\
      J & 0,16 & $1\cdot10^4$ & 5   & $2\cdot10^8$ & $1\cdot10^8$ & $5\cdot10^8$ & $4\cdot10^8$ & $2,5\cdot10^8$\\
      \cmidrule(r){2-9}
    \end{tabular}
  }
  \caption{Application characteristics and performance per offered solution}
  \label{tab:performance-of-offers}
\end{table}

    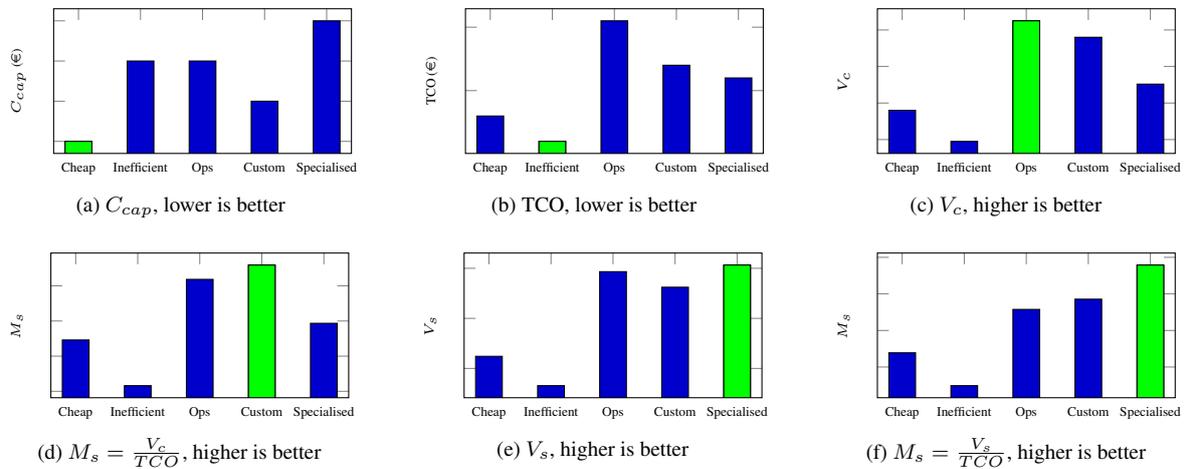
\begin{figure*}
      \centering
      \captionsetup{justification=centering}
      
      \subfloat[$C_{cap}$, lower is better\label{a}] {
        \pgfplotsset{
          width=5.5cm,
          height=3.5cm,
          compat=1.14,
          scaled y ticks=false,
          every tick label/.append style={font=\tiny},
          every axis/.append style={font=\tiny}
        }
        \begin{tikzpicture}
          \begin{axis}[
              symbolic x coords={Cheap, Inefficient, Ops, Custom, Specialised},
              xtick=data,
              yticklabel=\empty,
              ylabel=$C_{cap}$ (\euro),
            ]
            \addplot[ybar, fill=blue!80!black] coordinates {
              (Cheap, 250000)
              (Inefficient, 350000)
              (Ops, 350000)
              (Custom, 300000)
              (Specialised, 400000)
            };
            \addplot[ybar, fill=green] coordinates {
              (Cheap, 250000)
            };
          \end{axis}
        \end{tikzpicture}    
      }
      \hfill
      \subfloat[TCO, lower is better\label{b}] {
        \pgfplotsset{
          width=5.5cm,
          height=3.5cm,
          compat=1.14,
          scaled y ticks=false,
          every tick label/.append style={font=\tiny},
          every axis/.append style={font=\tiny}
        }
        \begin{tikzpicture}
          \begin{axis}[
              symbolic x coords={Cheap, Inefficient, Ops, Custom, Specialised},
              xtick=data,
              yticklabel=\empty,
              ylabel=TCO (\euro),
            ]
            \addplot[ybar, fill=blue!80!black] coordinates {
              (Cheap, 1300000)
              (Inefficient, 1100000)
              (Ops, 2050000)
              (Custom, 1700000)
              (Specialised, 1600000)
            };
            \addplot[ybar, fill=green] coordinates {
              (Inefficient, 1100000)
            };
          \end{axis}
        \end{tikzpicture}    
      }
      \hfill
      \subfloat[$V_c$, higher is better\label{c}]{
        \pgfplotsset{
          width=5.5cm,
          height=3.5cm,
          compat=1.14,
          scaled y ticks=false,
          every tick label/.append style={font=\tiny},
          every axis/.append style={font=\tiny}
        }
        \begin{tikzpicture}
          \begin{axis}[
              symbolic x coords={Cheap, Inefficient, Ops, Custom, Specialised},
              xtick=data,
              yticklabel=\empty,
              ylabel=$V_c$,
            ]
            \addplot[ybar, fill=blue!80!black] coordinates {
              (Cheap, 900000000)
              (Inefficient, 475000000)
              (Ops, 2125000000)
              (Custom, 1900000000)
              (Specialised, 1258750000)
            };
            \addplot[ybar, fill=green] coordinates {
              (Ops, 2125000000)
            };
          \end{axis}
        \end{tikzpicture}
      }
      
      \subfloat[$M_s=\frac{V_c}{TCO}$, higher is better\label{d}] {
        \pgfplotsset{
          width=5.5cm,
          height=3.5cm,
          compat=1.14,
          scaled y ticks=false,
          every tick label/.append style={font=\tiny},
          every axis/.append style={font=\tiny}
        }
        \begin{tikzpicture}
          \begin{axis}[
              symbolic x coords={Cheap, Inefficient, Ops, Custom, Specialised},
              xtick=data,
              yticklabel=\empty,
              ylabel=$M_s$,
            ]
            \addplot[ybar, fill=blue!80!black] coordinates {
              (Cheap, 692)
              (Inefficient, 432)
              (Ops, 1037)
              (Custom, 1118)
              (Specialised, 787)
            };
            \addplot[ybar, fill=green] coordinates {
              (Custom, 1118)
            };
          \end{axis}
        \end{tikzpicture}
      }
      \hfill
      \subfloat[$V_s$, higher is better\label{e}] {
        \pgfplotsset{
          width=5.5cm,
          height=3.5cm,
          compat=1.14,
          scaled y ticks=false,
          every tick label/.append style={font=\tiny},
          every axis/.append style={font=\tiny}
        }
        \begin{tikzpicture}
          \begin{axis}[
              symbolic x coords={Cheap, Inefficient, Ops, Custom, Specialised},
              xtick=data,
              yticklabel=\empty,
              ylabel=$V_s$,
            ]
            \addplot[ybar, fill=blue!80!black] coordinates {
              (Cheap, 621000)
              (Inefficient, 327750)
              (Ops, 1466250)
              (Custom, 1311000)
              (Specialised, 1531875)
            };
            \addplot[ybar, fill=green] coordinates {
              (Specialised, 1531875)
            };
          \end{axis}
        \end{tikzpicture}
      }
      \hfill
      \subfloat[$M_s = \frac{V_s}{TCO}$, higher is better\label{f}]{
        \pgfplotsset{
          width=5.5cm,
          height=3.5cm,
          compat=1.14,
          scaled y ticks=false,
          every tick label/.append style={font=\tiny},
          every axis/.append style={font=\tiny}
        }
        \begin{tikzpicture}
          \begin{axis}[
              symbolic x coords={Cheap, Inefficient, Ops, Custom, Specialised},
              xtick=data,
              yticklabel=\empty,
              ylabel near ticks,
              ylabel={$M_s$}, 
            ]
            \addplot[ybar, fill=blue!80!black] coordinates {
              (Cheap, 0.477692307692308)
              (Inefficient, 0.297954545454545)
              (Ops, 0.715243902439024)
              (Custom, 0.771176470588235)
              (Specialised, 0.957421875)
            };
            \addplot[ybar, fill=green] coordinates {
              (Specialised, 0.957421875)
            };
          \end{axis}
        \end{tikzpicture}
      }
      \caption{The offers evaluated against six cost and value measures. The superior offers for each measure are shown in green.}
      \label{fig:offer-evaluation}
    \end{figure*}
 

Each of these offers were evaluated using the model introduced in this paper, the results of which are shown in Figure \ref{fig:offer-evaluation}.
For each value measure, the superior solution is shown in green\footnote{All underlying data and analysis used in this paper are available here: \url{https://doi.org/10.5281/zenodo.2270842}.}.
While the offers are fictional and the use-case is obviously constructed, it is clear that, depending on the chosen selection criterion, a different solution wins, highlighting both the power and importance of the concept introduced in this paper.
More importantly, this example shows the dangers of selecting the wrong value measure for convenience or not carefully considering all possible components that make up the selected value measure.

In section \ref{sec:intro} we postulate that the useful (scientific) output of the system per invested Euro is the most useful value metric of a system.
Not using such a metric, and instead focusing solely on total cost of ownership, would, in this example, lead to the selection of the far inferior \textit{Inefficient} solution.

\section{Case studies}
\label{sec:case_studies}

To further illustrate the value of the conceptual model introduced in this paper, three radio astronomy use cases will be discussed: the use of the TITAN computer for one of the first operational radio interferometers, the LOFAR array and the SKA Science Data Processor.
We also highlight the variability of value over the lifetime of a compute system using the performance impact of a recent hardware vulnerability as an example.

\subsection{The TITAN Computer and the Mullard Radio Astronomy Observatory}
\label{sec:Titan}

The One-Mile Telescope, sited at the Mullard Radio Astronomy Observatory (MRAO) near Cambridge, was an early aperture synthesis telescope, and was the first designed to use the Earth rotation aperture synthesis technique.
It was conceived when the EDSAC II computer at Cambridge University was in operation, and was completed in 1964, as the TITAN computer came online.
TITAN was then used by the One-Mile, the Half-Mile and Interplanetary Scintillation Array (IPSA) telescopes, until TITAN was decommissioned in 1973. 
The One-Mile was explicitly designed to used the improved computing resources provided by TITAN, first to provide the control tapes for the telescope and then using the Fast Fourier Transform (FFT) to power the data analysis~\cite{1966MNRAS.134...87E}, ~\cite{1962MNRAS.125...39R}.
As Wilkes recalls: 
\begin{displayquote}
One day, Ryle came to me to say that he was planning the erection of a much larger telescope and to ask whether the Mathematical Laboratory could undertake to provide the computing support required.\cite{WilkesM.V.MauriceVincent1985Moac}, p.193
\end{displayquote}

TITAN was a ground-breaking computer itself, with hardware procured from the Ferranti company, and software developed by staff at the University of Cambridge, mostly from the Mathematical Laboratory.
The Mathematical Laboratory was, unusually for the time, already running as an effective computing service, where users applied for time with their projects (as is common with HPC resources today)\cite{1990Ahos}, \cite{WilkesM.V.MauriceVincent1985Moac}.
This differed from companies such as Ferranti and IBM, which were producing computers for the commercial market, and other universities, which were producing computers primarily as a way of investigating computers themselves (the purpose of instruments such as the Manchester Baby and CSIRAC), without explicit support for scientific research ~\cite{CroarkenMary1990Esci}. 

Although the University wished to buy a new computer to replace EDSAC II, it did not have a large capital budget available.
Thus they bought a heavily-discounted Ferranti Atlas (usual cost \pounds\,2 million; price actually paid: \pounds\,350,000 (approximately \pounds\,6-7 million today~\cite{currency})  with an additional \pounds\,75,000 for a large disk store ~\cite{CamComputing}).
However, the University now had to spend a lot of money on salaries to develop the software, but this cost was not explicitly tracked by the University, and their decisions were made purely on the $C_{cap}$.


The performance of TITAN, combined with David Wheeler's FFT algorithm~\cite{Ryle:1974}, allowed TITAN to do the calculations necessary for the first Earth-rotation aperture synthesis observations with the One-Mile telescope, and then to produce the first maps of the radio sky~\cite{1962MNRAS.125...39R}.
It was also used to support IPSA, which was used by Dame Jocelyn Bell Burnell to discover the first pulsar.

These scientific breakthroughs, backed by TITAN, won Tony Hewish and Sir Martin Ryle their joint Nobel prize for their innovative telescope design work~\cite{Ryle:65}. 
Furthermore, at least 30 PhD theses using the One-Mile, and the subsequent Half-Mile and IPSA, used TITAN-processed data, or used TITAN for theoretical modelling.\footnote{One of us (VA) checked the archived PhD theses of the Astrophysics Group, Cavendish Laboratory, University of Cambridge. TITAN may have been used for PhD theses in other groups; however, it is difficult to locate all of these 40 years later.}
It is unfortunately not possible to track all the papers that were produced with TITAN, as it was not comprehensively tracked at the time and not all authors note their use of TITAN.
Therefore there are aspects of TITAN's value that are not captured.

Nevertheless, TITAN delivered exceptional TVO extending well beyond its lifetime.
It was not only used in radio astronomy, although radio astronomy made unique use of its capabilities, but also in computer science (applications included one-way functions for storing passwords, timesharing systems, computer language research, early version control systems ~\cite{CamComputing}), crystallography (another field that used the FFT), statistics, Computer Aided Design, agronomy, and quantitative economic methods (for which one TITAN user, Sir Richard Stone, won the Nobel Prize for Economics)~\cite{Stone}\footnote{A copy of this work is held in the Library of the Department of Computer Science and Technology, University of Cambridge, classmark V75-14.})~\cite{Needham:1992}.

\begin{figure}
   \includegraphics[width=\columnwidth]{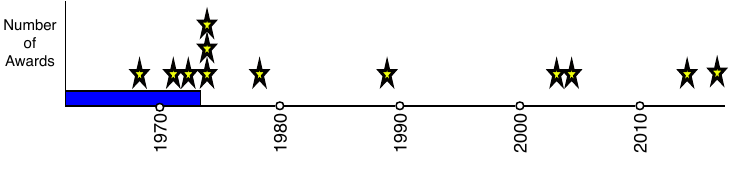}
   \caption{Awards given to TITAN radio astronomy users over time. The blue bar indicates when TITAN was active.}
   \label{fig:titan-awards}
\end{figure}

Many of the people who designed, programmed for, and used, TITAN were or became leaders of their fields, bringing rewards (both financial and reputational) to their institutions in the subsequent decades; thus TITAN provided a TVO that far outweighed cost of purchasing, developing, and running the system.
There is a significant ``long tail" to TITAN's value, exemplified by Dame Jocelyn Bell Burnell's receipt of the Royal Society Royal Medal in 2015, and the Special Breakthrough Prize in Physics (2018), both of which specifically cite her work on pulsars.
To illustrate the disparity between the lifetime of the TITAN system and its value, we have plotted major prizes won by TITAN users in radio astronomy, as compared to TITAN's lifespan, in Figure \ref{fig:titan-awards}.

Awards were not confined to the radio astronomy community. 
Eleven TITAN users have been elected Members of the Royal Society.\footnote{Frank Yates, Donald Lynden-Bell, David George Kendall, Maurice Wilkes, Sir Martin Ryle, Peter Swinnerton-Dyer, Malcolm Longair, Brian Pippard, Roger Needham, John Baldwin and Dame Jocelyn Bell Burnell ~\cite{RoyalFellows}.
``The Royal Society is a Fellowship of many of the world's most eminent scientists and is the oldest scientific academy in continuous existence'', and members must have made ``a substantial contribution to the improvement of natural knowledge, including mathematics, engineering science and medical science''~\cite{AboutRS,ElectRS}.}
TITAN users have won many other major awards in their fields, including: the Royal Statistical Society Guy Medal in Silver \footnote{Won by Georgle Kendall and MJR Healy}, the
Gold Medal of the Royal Astronomical Society\footnote{Won by Donald Lynden-Bell, Sir Martin Ryle, and Jeremiah P. Ostriker}, the London Mathematical Society De Morgan Medal \footnote{Won by D. G. Kendall}, the Faraday Medal \footnote{Won by Ryle, Maurice Wilkes, and Roger Needham}, the IEEE John von Neumann medal \footnote{Won by Wilkes} and the Karl G. Jansky Lectureship, which "is an honor established by the trustees of Associated Universities, Inc., to recognize outstanding contributions to the advancement of radio astronomy" \footnote{Awarded to Bernie Fanaroff and Dame Jocelyn Bell Burnell, both of whom used TITAN during their PhDs.}.
The precise role of TITAN in these awards is difficult to quantify; however, having TITAN available clearly provided important support and enablement for people at all stages of their careers --- precisely the purpose of a scientific computing resource.
Moreover, this lists only the very highest achievers amongst TITAN users; there are shallower network effects from the existence of TITAN, which are even harder to account for, but which indicate that resources such as TITAN are vital for the scientific community. 
Although a significant investment, both capital, as well as engineering and development, was required, TITAN's ten-year lifespan and high-impact and long-lasting contributions make its relative science value exceptional.
Even if the full list price of the hardware had been paid, the capital outlay would still have been justified by its scientific success, which far outshone other contemporaneous systems.

\subsection{LOFAR}
\label{sec:lofar}
LOFAR, the LOw Frequency ARray~\cite{haarlem:2013}, is a modern low-frequency large-scale distributed radio telescope in the Netherlands, with international stations in various European countries.
The concept and design of the LOFAR telescope, which started in the late 1990s, is a study in trading off value and cost.
A number of early papers discussing the telescope concept and initial design~\cite{Bregman:1999,Bregman:2000}, as well as some retrospective analysis of the design considerations~\cite{Bregman:2012}, make this a particularly interesting instrument to study.

As discussed above, modern radio interferometry was made possible by the availability of abundant and affordable compute resources.
In LOFAR, this concept is taken even further by replacing a small number of large parabolic reflectors with many simple, cheap and omni-directional dipole antennas and software-based digital beamforming.
Essentially, many simple antennas are combined, by coherent addition, into a single virtual receiver.
Early design concepts for this low-frequency array that could act both as a technology demonstrator for the future Square Kilometre Array, as well as scientifically open a relatively unexplored frequency range, identify a ``processing window of opportunity''.
This early concept predicted that, while computational cost for the processing required for this low-frequency array was at the time infeasibly large, it would become affordable, assuming Moore's law continued to apply, after 2003.

In further work, instrument sensitivity was defined as the key value parameter (and thus a measure for the TVO of the instrument) for the design trade-offs in this instrument~\cite{Bregman:2004}, although other value measures such as survey speed and resolution were also taken into account.
In order to achieve optimal performance over cost, all main constituents of the complete LOFAR system were designed to have a similar marginal performance over cost ratio.

This analysis shows that both TVO and TCO for the LOFAR telescope in general, and the digital processing systems in particular, were carefully considered early on in the conceptual design phase of the instrument.
A clear choice was made to use sensitivity over other technical or scientific metrics, such as survey speed or resolution, as a measure for the total value of the instrument.
We note that this implicitly assumes this technical measure translates to scientific value.
Regardless of this technical measure, suitability for a small number of key science projects was also a key design consideration in the development of LOFAR.
Furthermore, the cost of the digital processing system was analysed, and, more importantly, judged to become affordable at some point in the mid future.
This realisation allowed development of the instrument, and its associated software infrastructure, to start before the required compute capacity became financially feasible.

Since its opening in June 2010, one measure of LOFAR's science value, the number of peer-reviewed scientific publications using LOFAR produced data, has been monitored~\footnote{\url{https://old.astron.nl/radio-observatory/lofar-science/lofar-papers/lofar-papers}}.
A different way to express the value, or in this case more accurately the return on investment of a science instrument, is to evaluate how much of the invested money is reinvested in the local (national) economy.
A Dutch research institute that specialises in research on the impact of science, Rathenau, recently studied the LOFAR telescope and the Dutch contribution to three other major science instruments: CERN~\footnote{\url{https://home.cern/}}, ESRF~\footnote{\url{https://www.esrf.eu/}} and ITER~\footnote{\url{https://www.iter.org/}}.
They defined a return coefficient ($R$) as the capital reinvested in the national economy, divided by the Dutch contribution in that instrument.
The results, published in Dutch~\cite{rathenau:2019}, are summarised in table~\ref{tab:return-coefficients}.
While it is difficult to compare four completely different instruments, this work shows that the financial return of the LOFAR telescope for the Dutch economy has been excellent.
This is due to the fact that most, if not all, of the IP was developed in the Netherlands, and therefore production of those components, even for international stations, is likely to occur there as well.

\begin{table*}
  \centering
  \begin{tabular}{lcccccc}
    \toprule
                                & CERN       & ESRF       & \multicolumn{2}{c}{ITER}    & \multicolumn{2}{c}{LOFAR}  \\
                                & average    & average    & 2008-2017    & 2008-2015    & 2004-2013    & 2014-2017 \\
                                & per year   & per year   & construction & construction & construction & operations \\
                                &            &            & incl. grants & grants only  &              & per year \\
    \cmidrule(r){2-7}
    Total investment            & 1,104M CHF & 90M \euro  & 6,120M \euro & 4,581M \euro & 92M \euro    & 4.3M \euro \\
    Total Dutch investment      & 50,9M  CHF & 2.7M \euro & 161M \euro   & 120M \euro   & 81.2M \euro  & 3.4M \euro \\
    Dutch contribution          & 4.61\%     & 2.97\%     & 2.63\%       & 2.63\%       & 88.3\%       & 77.8\% \\
    \cmidrule(r){2-7}
    Total expenditure           & 343M CHF   & 57M \euro  & 4,330M \euro & 101M \euro   & 92 \euro     & 4.3M \euro \\
    To the Netherlands          & 8.7M CHF   & 0.58M \euro& 7.9M \euro   & 4.3M \euro   & 89.2 \euro   & 4.1M \euro \\
    Dutch ROI                   & 2.54\%     & 1.01\%     & 0.07\%       & 4.18\%       & 97\%         & 96.5\%    \\
    \cmidrule(r){2-7}
    \textbf{Return coefficient} & 0.55       & 0.34       & 0.07         & 1.59         & 1.10         & 1.24\\

    \bottomrule
    
  \end{tabular}
  \caption{Return coefficients for the Dutch economy for four large scale science infrastructure projects (source, Rathenau institute~\cite{rathenau:2019})}
  \label{tab:return-coefficients}
 \end{table*}

The hierarchical and modular nature of the LOFAR system has allowed several dedicated systems to be added to the telescope to increase its scientific value at modest cost.
While some, like Dragnet (described in section \ref{sec:dragnet}), were just plug-in systems that required little to no additional engineering to add to LOFAR, others, like AARTFAAC (see section \ref{sec:aartfaac}), require raw antenna data not available in standard LOFAR observation modes.
We will explore the cost and value considerations of some of these components in the following sections.

\subsubsection{LOFAR correlator and beamformer systems}

A key signal processing component of the instrument, the LOFAR correlator and beamformer, and specifically its hardware evolution, is relatively well described.
This part of the instrument is algorithmically simple and the required functionality is fairly constant.
Therefore, for this specific example, cost (with all its different components), operational availability, and lifespan mostly determine the relative science value of the correlator and beamformer.
Early concepts for the LOFAR central processor show a 1600 node hybrid cluster compute system that uses conventional processors and dataflow co-processors to process the data~\cite{deVos:2001, Schaaf:2003}.
While feasible, the considerable size of this compute concept meant that a bespoke supercomputer was a viable and, more importantly, cost-effective alternative.
In 2003, an IBM Blue Gene/L, briefly the fastest supercomputer in Europe, was installed as the central correlator and beamformer for LOFAR~\cite{Romein:06}.
This was upgraded to a much smaller, IBM Blue Gene/P in 2008~\cite{Romein:10}, that was not only more powerful, but also considerably more energy efficient.
Whereas the total lifetime computational and scientific value of this new system was similar, its reduced operational costs, as well as improved software environment made its relative science value considerably higher than the previous Blue Gene/L.
However, supercomputers are inherently expensive, so research into more cost-effective solutions continued~\cite{Schaaf:2004,Nieuwpoort:10}.
This eventually resulted in the procurement and commissioning of a much smaller and more affordable GPU-based correlator and beamformer platform, Cobalt~\cite{Broekema:18a}.
A more capable second generation of this system, Cobalt 2.0, started operations in 2019~\footnote{\url{https://old.astron.nl/cobalt20-sets-stage-fully-multitasking-lofar}}.

The timeline of the LOFAR correlator and the construction of the instrument as a whole is interesting to study.
As mentioned above, the telescope was opened in 2010 and at that time the initial Blue Gene/L correlator and beamformer has already been replaced.
While it is not accurate to say Blue Gene/L was never used in production as the LOFAR correlator and beamformer, it is clear that it was procured and installed early.
Arguably, its cost was considerable (although the actual investment was never made public), and its value limited.
However, the strategic alliance and collaboration agreement between ASTRON and IBM was an important consideration in securing sufficient construction funding for LOFAR.
Furthermore, spare computational capacity was made available to other scientific users.
Therefore, while the total lifetime scientific value of the Blue Gene/L correlator for the LOFAR telescope was low, its general value for the LOFAR telescope was extremely high and its total lifetime scientific value for the wider community was comparable to other high-performance computing systems.
Nevertheless, the Blue Gene/L system was never used to its full potential in the LOFAR telescope, and even the Blue Gene/P system was significantly under-utilised for most of its lifetime.
These systems did however provide extremely valuable experience that was essential to the success of Cobalt and was used to great effect in the hardware design of the SKA Science Data Processor.
Whether this was worth the significant initial capital investment is beyond the scope of this paper.

When the LOFAR Blue Gene/P was nearing the end of its service life, a feasibility study into possible upgrades was undertaken~\cite{holties:2012}.
Four drop-in replacement options (Blue Gene/Q, an FPGA-based Uniboard system, a CPU-based cluster with GPU accelerators, and a CPU-based cluster) were evaluated for risk, development effort, cost, power consumption and scalability.
It is clear from these selected criteria that various cost components were carefully considered, while value was expected to be equal among the contenders considering any new system was expected to replicated the functionality of the existing Blue Gene/P based correlator and beamformer.
A cluster with GPU accelerators was judged to be the most cost-effective solution, based on low cost and power consumption, good scalability, and relatively little development effort required.
By extension this was therefore also the option with the highest relative science value, and selected for implementation as the Cobalt correlator and beamformer~\cite{Broekema:18a}.



\subsubsection{AARTFAAC}
\label{sec:aartfaac}
While the LOFAR correlator and beamformer described above are integral parts of the original LOFAR design, AARTFAAC is an add-on system that was added to increase functionality and enable additional science cases.
The Amsterdam–ASTRON Radio Transients Facility and Analysis Center (AARTFAAC) system~\cite{prasad:2016} is a real-time all-sky transient detection system.
Data from a subset of LOFAR antennas is duplicated during normal LOFAR operations and processed independently into all-sky images of the low-frequency radio sky that can subsequently be monitored for bright transient events.
This is a significant advance over the capabilities of the original LOFAR system, which was only possible due to investments made early in the LOFAR project to over-provision both the bandwidth of the LOFAR station ring network and the LOFAR Wide Area Network.
For this specific addition, a custom shim was added the station data transport ring: Uniboard-RSP Interface boards.
These duplicate raw antenna data, normally beamformed in RSP boards, to AARTFAAC Uniboards.
Furthermore, the FPGA firmware on the LOFAR core stations that take part in the AARTFAAC system had to be modified to generate the additional AARTFAAC packets.
Data from the AARTFAAC system is transported to dedicated processing nodes located in the same central processing facility as the LOFAR correlator and beamformer, sharing spare network capacity.

While AARTFAAC adds undoubtable value to the LOFAR telescope, its addition required significant additional engineering and manufacturing.
In particular the additional firmware programming requires the use of scarce resources that are generally overcommitted.
We will not discuss the relative merits of this addition over others, or whether the investment was valuable or not.
However, we do note that additional investment in the development of data spigots at the LOFAR station during construction would have made the development of AARTFAAC much cheaper and easier.
This was considered during design, but technology had not progressed sufficiently; the additional cost would have been significant and the idea was shelved.

  

\subsubsection{DRAGNET}
\label{sec:dragnet}
Whereas AARTFAAC is a real-time transient monitor that operates in UV-space, the DRAGNET cluster~\cite{bassa:2017} is a non-real-time pulsar and transient search system that operates in the time domain.
It takes beamformed data from the Cobalt correlator and beamformer and uses blind coherent de-despersion to identify fast transients and millisecond pulsars.
This system has demonstrated its value by the discovery of the second fastest-spinning pulsar to date, and one of the first at such low observing frequencies~\cite{bassa:2017a}.

The DRAGNET system consists of 23 nodes, each of which has 4 NVIDIA Titan X GPUs that provide the bulk of the processing capacity.
Its source data is produced by the LOFAR Correlator and Beamformer, Cobalt.
Data is stored locally and processed non-real-time, resulting in a pulsar and/or transient candidate list for further analysis.

Since DRAGNET uses a standard LOFAR data product as input, only limited modifications were necessary to integrate the system into the LOFAR telescope.
The only major investment, apart from the cluster and dedicated software for DRAGNET itself, was the integration of the system into the LOFAR monitoring and control system.
DRAGNET has added significant capability to the LOFAR telescope: the ability to search for extremely fast-spinning pulsars, and a way to detect fast transient events that would be missed by the original LOFAR telescope.
This adds significant additional value to the instrument, since it allows new science cases to be explored.
The majority of the additional investment was in the actual cluster and the software development needed to process the data, with limited investment needed to modify the existing system.



\subsubsection{International LOFAR stations}

International LOFAR stations are not just valuable parts of the International LOFAR Telescope (ILT), these can also operate in \textit{local} or \textit{standalone} mode.
In this mode, station data is not sent to the central LOFAR correlator and beamformer, but instead redirected to a local system and can thus act as a fully functional telescope in its own right.
The comparatively small size of these stations, and the low observation frequency, make them relatively unsuited for imaging observations, so most effort has gone into local transient and pulsar search.
The ARTEMIS backend~\cite{armour:2012} was developed as a real-time GPU accelerated suite of software to search for these events in data from modern radio telescopes.
Four international stations are equipped with such systems~\cite{serylak2012}.

Changing an international LOFAR station to stand-alone mode is, from a high level, as easy as changing destination IP number and MAC address of the receiving nodes.
The ability to use these international stations in this mode can be partly attributed to the extensive use of standardised protocols and interfaces, as well as the modular nature of the LOFAR telescope.
This means that LOFAR is potentially a large collection of independent instruments.

One international station, the French station near Nan\c{c}ay, differs significantly from any other antenna field in the LOFAR instrument.
Apart from the  low- and high-band antennas as in every international station, an unused third analog data path in the LOFAR station hardware is used to add a cluster of 96 mini-arrays, each of which consists of 19 antennas sensitive from 10 to 87\,MHz~\cite{zarka:2015}.
The resulting giant extension of LOFAR, NenuFAR, while not as large as the LOFAR telescope, adds a similar number of low-band antennas to the instrument as all other stations combined (1938 vs $\sim$2700).
In stand-alone mode, NenuFAR, currently under construction and accepting early science proposals\footnote{\url{https://nenufar.obs-nancay.fr/en/astronomer/}}, intends to support a wide range of data products, very similar to those produced by the LOFAR telescope.
This shows that NenuFAR is a powerful instrument itself, especially for pulsar and radio transient science.
A dedicated correlator and beamformer, based on the newly commissioned Cobalt 2.0 correlator and beamformer, is currently being installed.

This extension to the French international station was made possible by the availability of an unused analog data path in the LOFAR station hardware.
This data path was intended for a third receiver type, eliminated early in the design process for cost reasons.
In Dutch LOFAR stations this data path is used to connect half of the low-band antennas.

Finally, a LOFAR station was constructed in Lapland, Finland, near Kilpisj\"arvi, well above the arctic circle~\cite{mckay:2014}.
This station, KAIRA, is not part of the International LOFAR Telescope (ILT) and not connected to the rest of the LOFAR network.
Instead it is used exclusively in stand-alone mode, primarily for atmospheric imaging using reflected transmissions from a number of remote radar sites.
Experiments have shown fringes on recorded data between it and the international LOFAR station at Effelsberg in Germany, proving that for exceptional experiments it is possible to add the station to the LOFAR array, albeit not in real-time.


\subsubsection{Retrospective}
The LOFAR concept design identified a period in time where the relatively high impact of ground-breaking radio astronomical research in a relatively unexplored frequency range, combined with dropping costs for computing, would result in an instrument with optimal relative science value.
During its design and operational lifetime, the LOFAR correlator and beamformer in particular has benefited from continued development of cost-optimised solutions to improve the relative science value of an already successful and cost-effective instrument.
The modular nature of the LOFAR telescope enabled the addition of additional systems to the instrument, further increasing its science value.

We note that the cost and value analysis of these additional systems was not as rigorous as that done for the original LOFAR system.
While the engineering challenges of such add-on system were generally considered, the operational impact was often under-estimated and  (un)availability of critical development resources lead to significant slippage in project schedules.
Within ASTRON this has led to a more formal and structured application process for funding and the adoption of more rigorous systems engineering practices.
For the second phase of LOFAR we are considering the establishment of a LOFAR architecture team to centralise and formalise the responsibility for the considerations on cost and value for the instrument.
Modern distributed radio telescopes are, due to their inherent modular nature, exceptionally adaptable and extendable.
Taking possible extensions into account during the design of a new instrument will make the addition of such extensions easier and thus cheaper.

When looking at radio telescope systems as a whole, instead of just the compute systems they rely on, scientific value is the better understood factor while the sum of all costs is often not fully appreciated.
This is, at least to some degree, a result of the funding model for scientific instruments.
Funding proposals are evaluated on scientific merit first, and cost second.
Furthermore, costing an addition to a complex distributed sensor system, like the LOFAR telescope, is exceedingly complex and prone to overseeing non-trivial component costs.

\subsection{Spectre and Meltdown: how value of an existing system may change unexpectedly}
We argue in this paper that we can try to estimate the total lifetime computational value of a hardware system beforehand.
However, value is not constant over time and may be impacted by external factors beyond the control of the user.
In January 2018 a number of critical and widespread flaws in the hardware design of current generation processors were published~\cite{Lipp2018meltdown,Kocher2018spectre}.
These unparalleled hardware vulnerabilities hit virtually every installed compute system currently in operation.
While many software bugs may cause temporary performance issues, or cause delays in achieving top performance, the mitigations implemented to address these unprecedented flaws in processor design caused a completely unexpected and major reduction in performance of current systems, including otherwise well-performing systems. 
Due to the nature of these flaws, critical separation failures in performance-critical speculative execution, mitigation efforts in processor microcode and operating system kernel, have resulted in significant performance impacts, thus reducing the value of existing compute systems.
In particular I/O heavy workloads, such as those encountered in the LOFAR correlator and beamformer, that cause large numbers of context switches are expected to see performance reduced by very significant amounts~\cite{KPTI-performance}.
For the Linux kernel, the dominant operating system in both high-performance computing, as well as distributed computing applications, these are known as Kernel Page Table Isolation (KPTI).
These are kernel level fixes, that can be activated or deactivated at boot-time with a kernel boot parameter.

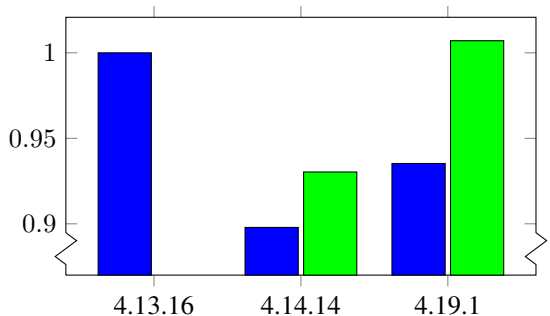
\begin{figure}
  \begin{tikzpicture}
    \begin{axis}[
        width=\columnwidth,
        height=5cm,
        axis y discontinuity=crunch,
        ybar,
        ymin = 0.87,
        enlarge x limits=0.3,
        bar width=7mm,
        xtick=data,
        symbolic x coords = {4.13.16,4.14.14,4.19.1},
      ]
      \addplot [fill=blue] coordinates {(4.13.16,1.0) (4.14.14,0.897891525276834) (4.19.1,0.935286271634661)};
      \addplot [fill=green] coordinates {(4.14.14,0.930302466465037) (4.19.1,1.00708543824084)};
    \end{axis}
  \end{tikzpicture}
  \caption{Maximum UDP/IP packet receive performance for three kernels, normalised to the oldest kernel. Blue shows the default configuration, green when Spectre and Meltdown v2 mitigations are turned off.}
  \label{fig:spectre-patch-impact}
\end{figure}

We illustrate the performance impact of these mitigating efforts in Figure \ref{fig:spectre-patch-impact}.
We test three Linux kernels, one released just before the announcements mentioned above (4.13.16), one that includes the initial mitigating patches (4.14.14) and one more recent kernel (4.19.1) in which the mitigations have been in place for some months.
Since a key task in the correlator and beamformer systems in LOFAR involves receiving large numbers of UDP/IP streams, we measure performance impact, and therefore the hit on value, by trying to receive as many UDP/IP packets as possible on a CPU-bound system with a 40~GbE device.
Results are normalised to the performance of the oldest kernel, which, for reference, achieved around 1,65 million packets per second.

This measurement shows that the value of a system has the potential to change over time (here between 5\% and 10\%), and may be affected by factors and risks outside its operators' and designers' control.
In this particular case, most of the performance impact may be avoided by turning off page table isolation (\verb!nopti!) and retpoline (\verb!nospectre_v2!) at boot time, at the cost of accepting that the system is trivially exploitable (which may be acceptable for a dedicated cluster behind a firewall).

\subsection{SKA SDP}
The Square Kilometre Array (SKA) is a next-generation radio telescope, currently in the design phase.
It will consist of two telescopes, a low-frequency telescope (50-350MHz) consisting of ~130,000 antennas in over 500 stations in Western Australia, and 133 mid-frequency dishes (4350MHz-14GHz) in South Africa, which latter will be joined to the existing MeerKAT telescope.
These telescopes will be constructed and deployed in phases across a 5-year period.
The SKA is designed to achieve exceptional scientific value, and to enable potential Nobel Prize-winning research~\cite{SKA}.

A key component of this instrument is the Science Data Processor (SDP), where instrument data, produced by specialised correlator hardware, is turned into science-ready data products, such as radio astronomy images, using high performance general-purpose compute systems~\cite{alexander:2019a}.
There will be data centres in Perth and Cape Town, which will host an SDP for each country, where the data will be received, processed, and transmitted for use by astronomers. 
The data rates from SKA will be extremely large: each telescope will output up to 3.1Tb/s from the correlator.
The main function of the SDP is thus to perform a data reduction, outputting data products that are able to be used by scientists, but which are also somewhat easier (and cheaper) to store and transmit.
The (in)ability of the SDP to perform this function may impact the science that can be performed by the SKA as a whole: if it takes too long to reduce the data, or the SDP cannot reduce the data by a sufficient factor, less data- or compute-intensive observations will have to be scheduled~\cite{alexander:2019a}.
Thus the design of the SDP is critical to the scientific value of the telescope.

In order to maximise its relative science value, the SDP will use a mix of custom-designed software components and off-the-shelf software.
In order to reduce TCO, the SDP will make extensive use of existing technologies.
A platform management system is envisioned to provision and organise its compute resources.
Such a system allows for the automation of compute deployment, at the cost of a mild computational overhead.
This saves on operator time, and allows for reliable and reproducible deployment of operating systems and other support services. 
The reduced operator time needed and increased reliability drive down operational costs ($C_{ops}$); reproducibility renders it easier and quicker (hence cheaper) to detect bugs.  
OpenStack, an open source platform management product in use by HPCs and data centres, is a candidate solution for this, in part because SKA is already working with CERN on improving OpenStack technologies.
The SKA will save cost of development ($C_{dev}$) and ongoing maintenance costs by using this off-the-shelf open source software, rather than writing their own suite of complicated software for the same purpose.
The viability of this approach has already been prototyped ~\cite{taylor:2019}.

In addition, in order to improve TVO, a new suite of astronomy data processing software will be developed, focusing on a highly reusable modularised architecture. 
The principle idea is to create low-level software modules that can be reused by many data processing pipelines~\cite{alexander:2019b}.
However, rather than using existing code from existing telescopes, these modules will be newly implemented for two reasons: scalability in parallel environments and maintainability over the expected 50-year telescope lifetime.


Providing modules that can easily be connected for use in large clusters is key for the SDP, as, without taking advantage of the inherent parallelism available in a lot of astronomy data processing, it will be difficult to achieve the data throughput necessary.
This modularisation not only allows designing for an embarrassingly parallel processing environment, it also permits programmers to quickly and easily provide new modules for optimised use with new hardware, and implement new algorithms for new science without rewriting other parts of the software infrastructure.
This is explicitly to reduce $C_{dev}$, by anticipating the need to port code to new, potentially very different, architectures, in contrast to the issues LOFAR experienced, as noted in \ref{sec:lofar}.
This also allows the SDP to run on generic COTS hardware, while also allowing for future software-hardware co-design for key algorithmic components. 
Similarly, modularisation of code handling hardware interfaces allows for pivoting to new technology --- an inevitability in a long-lived project. 

Maintainability is also a key driver for writing new code: technical debt accrues in software projects over time, as programmers can end up prioritising writing code quickly, rather than writing it well, or with an eye to help reduce maintenance costs.
Some of the commonly-used radio astronomy code, such as CASA, has parts that are nearly 40 years old, and which were not designed to be used in highly parallel compute systems. 
Thus the SKA has the opportunity to reduce its total lifetime costs by investing in new code that is designed to be more easily maintainable, especially around providing new algorithms and pipelines for its highly parallel environment.
Proofs of concept of this approach have similarly been prototyped, to verify ease-of-use and explore the scalability required for SKA~\cite{Allan:2019,Cornwell:2019}. 

This requires a significant up-front investment for rewriting code -- SKA SDP software accounts for approximately 8.2\% of the SKA construction budget, compared to 7.1\% for the VLT, 5.7\% for ALMA and 4.3\% for ASKAP~\cite{kemball:2004,guzman:2014}.
We note that for SKA SDP this is processing software only, excluding telescope manager -- functionality that is included in the figures for VLT, ALMA and ASKAP.
However, this will improve TVO, by making it easier to make efficient use of the data processing hardware, by making it easy to implement new algorithms, and by isolating where code changes to support those algorithms are needed. 
This should thus reduce some of the maintenance cost of the SDP, and improve its ability to unlock new science across the lifespan of the telescope, albeit at an increase in upfront development cost ($C_{dev}$). 
The SDP will also undertake a phased hardware deployment, to provide compute when it is needed to support the increasing number of antennas and dishes on the ground, which will both keep capital costs lower, and reduce overall operational costs ($C_{ops}$).
Furthermore, the deployment of hardware later on in the project allows hardware to be tailored to the software and vice versa, similar to the Cobalt correlator and beamformer in LOFAR, improving total relative science value of the resulting system.
The SDP is for the SKA thus deliberately considering and trading off in different areas, the TCO and TVO of the system, with some decisions made to manage cost, and others to maximise total lifetime scientific value.

\section{Related work}
\label{sec:related}
This work is a form of hardware-software co-design, as practised in the design of compute systems for large-scale science instruments.
However, up to now, hardware-software co-design has focused mostly on more easily measured metrics, such as cost, power consumption and peak performance.
Furthermore, while the literature often speaks of the importance of application co-design, the metrics used are agnostic and described mostly in terms of cost functions and constraints in energy and capital.
In this paper we explore what these systems are really built for, and what a suitable measure for their performance would be.

This work can be considered a specialisation of general cost-benefit analysis in economics.
Whereas cost-benefit analysis normally evaluates the social or financial benefit of a certain investment, this paper focuses on the scientific benefit in particular.
There is research that introduces the concept of total value of ownership~\cite{wouters:2005} in accounting, but this is introduced as potential future research as an extension to TCO based decision making and not expanded upon.
In that paper it is claimed that TVO builds on the concept of value as described in marketing literature.

Total value of ownership, also referred to as total value of opportunity, is also a metrics-based methodology for measuring and analysing the business value of enterprise IT investments~\cite{apfel:2003}.
This is an extension of TCO analysis, where both cost and any benefits of the proposed investment, tangible or intangible, are considered.

Value Engineering, Value Management and Value Analysis in Systems Engineering describe processes to achieve an optimal solution~\cite{incose:2015}.
This optimal solution is based on stakeholder value metrics; the processes are agnostic to these.
In this paper we take the stakeholder view, describing and enumerating the value metric, while not considering the detailed processes required to optimise these.

Some work was done to analyse the societal impact of the High-Luminosity Large Hadron Collider (HL-LHC) upgrade of the LHC~\cite{florio:2016, bastianin:2018}, predicting a larger than 90\% chance of positive net economic benefit to society based on Monte Carlo simulations.
These simulations estimate the economic returns from diverse benefits such as value of training for students, technological and industrial spillover, cultural effects for the public and academic publications.
A comparison of the impact of the upgrade to the LHC with the non-upgraded instrument was also presented.
An impressive effort is made to estimate the total cost of the current LHC, a difficult task even though all CERN expenses are well documented, due to the many in-kind contributions by member and non-member states.
Societal impact analysis are very useful for funding agencies to gauge the value of an instrument to society using \textit{scientific} and \textit{objective} criteria.
However, analysis of the methodology found many ambiguities and the scientific benefits of the LHC is given as less than 2\% of the total impact of the instrument using this method: a drastic underestimation~\cite{schopper:2016}.
Furthermore, it was found that the societal impact of CERN's mission, ``promote science and bring nations together'',  was impossible to measure, since no way has been developed to measure in economic terms the success of the second objective.
In comparison , the concepts introduced in this paper look at more immediate impact, computational or scientific, and attempt to be more directly useful when making design choices.

Recent work on design optimisation of low-frequency telescopes using cost constraints~\cite{boonstra:2018} takes a slightly different and more domain specific approach.
Here, an attempt is made to model both cost and scientific performance of a radio telescope using Lagrange multipliers.
Scientific performance, defined by two instrumental figures of merit -- sensitivity and survey speed, is optimised using both models and an assumed fixed capital budget.
The LOFAR architecture as build, and the SKA phase 1 baseline design are analysed using the introduced model and variants optimised for survey speed and sensitivity are proposed.
This methodology focuses on receiver and front-end optimisation and mostly ignores the cost required for compute capacity or how this scales with the number of stations and length of baselines.
While the cost model does include a central correlator and beamformer, its model is exceedingly simple.
Calibration, imaging and other post-processing costs, as well as long-term storage of data products, monitoring and control and operational costs are not modelled.
Furthermore, we note that the chosen degrees of freedom in this paper, number of stations and number of antennas, have an enormous impact on required compute capacity for calibration and imaging.
In this paper we take a more generic approach that is not limited to radio astronomy and that focuses on the cost and value of the compute systems that are not considered by Boonstra et al.

\section{Summary and conclusions}
\label{sec:conclusions}
In this paper we introduced a more formal way to reason about cost and value of compute resources, both hardware and software.
We suggested that a focus on minimising cost alone is not sufficient to design an optimal solution.
The introduction of several new concepts, total value of ownership, total lifetime computational value, total lifetime scientific value and relative science value, gives us the \emph{vocabulary} to effectively discuss routes towards more optimal solutions.
Although both total lifetime computational value and especially total lifetime scientific value are difficult to quantify, and we do not expect anyone to do so using the formulas given in this paper, we do show a number of components that allow us to reason effectively about this metric.

We provided a number of case studies in which we demonstrate the concepts introduced in this paper. 
We can see the utility of explicitly considering a metric of total lifetime scientific value, as the TITAN computer sought only to minimise capital cost (which happily led it to deliver truly exceptional value), whereas the SKA designers are explicitly allowing for relatively high costs in some areas to maximise total scientific value.
In the LOFAR use case we noted the explicit trade-off made between high-impact science and dropping cost for computing, which led to an identified ``processing window of opportunity'' some years in the future where relative science value was perceived to be optimal.
Some of the later additions to LOFAR were discussed, each adding their own value to the complex machinery that is the LOFAR telescope.
Finally we showed, using a recent highly publicised processor flaw and its mitigating patches, that the total computational value of a system may potentially change over a system's lifetime. 
Together, these concepts and case studies provide a framework for decision makers, principal investigators, designers, and engineers of computing solutions to reason about the optimal solutions, in hardware or software, for their applications.




\section*{Acknowledgements}
The authors would like to thank Yan Grange, \'{A}gnes Mika, Bram Veenboer and Cees Bassa at ASTRON for their valuable feedback on early versions of this manuscript.
We would like to thank Dr Elizabeth Waldram and Professor Malcolm Longair from the Cavendish Laboratory, Cambridge, for their help with understanding the impact of the TITAN Computer, and the Librarian of the Department of Computer Science Library at the University of Cambridge for access to the TITAN archival material.

\bibliographystyle{plain}
\bibliography{escience}

\begin{thebibliography}{10}

\bibitem{CamComputing}
Haroon Ahmed.
\newblock {\em Cambridge Computing: The First 75 Years}.
\newblock Third Millennium Publishing Limited, London, UK, 2013.

\bibitem{alexander:2019b}
P.~Alexander, V.~Allan, U.~Badenhorst, C.~Broekema, T.~Cornwell, S.~Gounden,
  F.~Graser, K.~Kirkham, B.~Mort, R.~Nijboer, B.~Nikolic, R.~Simmonds,
  J.~Taylor, A.~Wicenec, and P.Wortmann.
\newblock {SDP System Module Decomposition and Dependency View}.
\newblock Technical report, SDP Consortium, 2019.

\bibitem{alexander:2019a}
P.~Alexander, B.~Nikolic, V.~Allan, C.~Broekema, M.~Deegan, and P.Wortmann.
\newblock {SKA1 SDP High Level Overview}.
\newblock Technical report, SDP Consortium, 2019.

\bibitem{alexander:2013}
Paul Alexander, Chris Broekema, Simon Ratcliffe, Rosie Bolton, and Bojan
  Nikolic.
\newblock {SDP Element concept}.
\newblock Technical report, SDP Consortium, 2013.

\bibitem{Allan:2019}
V.~Allan, B.~Nikolic, M.~Farreras, T.~Cornwell, and R.~Lyon.
\newblock {SKA1 SDP Execution Frameworks Prototyping Report}.
\newblock Technical report, SDP Consortium, 2019.

\bibitem{apfel:2003}
Audrey Apfel and Michael Smith.
\newblock {TVO methodology: Valuing IT investments via the Gartner business
  performance framework}.
\newblock {\em Strategic Analysis Report}, 2003.

\bibitem{armour:2012}
Wesley Armour, Aris Karastergiou, Michael Giles, Chris Williams, Alessio Magro,
  Kimon Zagkouris, Sarah Roberts, Stefano Salvini, Fred Dulwich, and Ben Mort.
\newblock {A GPU-based survey for millisecond radio transients using ARTEMIS}.
\newblock {\em ASP Conference Series}, 461, Astronomical Data Analysis Software
  and Systems XXI:33 -- 36, 2012.

\bibitem{Bal:2016}
H.~Bal, D.~Epema, C.~de~Laat, R.~van Nieuwpoort, J.~Romein, F.~Seinstra,
  C.~Snoek, and H.~Wijshoff.
\newblock A medium-scale distributed system for computer science research:
  Infrastructure for the long term.
\newblock {\em Computer}, 49(5):54--63, May 2016.

\bibitem{bassa:2017}
C.~G. {Bassa}, Z.~{Pleunis}, and J.~W.~T. {Hessels}.
\newblock {Enabling pulsar and fast transient searches using coherent
  dedispersion}.
\newblock {\em Astronomy and Computing}, 18:40--46, Jan 2017.

\bibitem{bassa:2017a}
C.~G. {Bassa}, Z.~{Pleunis}, J.~W.~T. {Hessels}, E.~C. {Ferrara}, R.~P.
  {Breton}, N.~V. {Gusinskaia}, V.~I. {Kondratiev}, S.~{Sanidas}, L.~{Nieder},
  C.~J. {Clark}, T.~{Li}, A.~S. {van Amesfoort}, T.~H. {Burnett}, F.~{Camilo},
  P.~F. {Michelson}, S.~M. {Ransom}, P.~S. {Ray}, and K.~{Wood}.
\newblock {LOFAR Discovery of the Fastest-spinning Millisecond Pulsar in the
  Galactic Field}.
\newblock {\em Astrophysical Journal, Letters}, 846(2):L20, Sep 2017.

\bibitem{bastianin:2018}
Andrea Bastianin and Massimo Florio.
\newblock {Social Cost-Benefit Analysis of HL-LHC}, 2018.

\bibitem{boonstra:2018}
Albert-Jan Boonstra and Ronald Nijboer.
\newblock {Radio Telescope Design Optimization Using Costing Constraints:
  Extrapolating LOFAR Costing to the Square Kilometre Array}.
\newblock {\em Radio Science}, November 2018.

\bibitem{Bregman:1999}
Jaap~D. Bregman.
\newblock Design concepts for a sky noise limited low frequency array.
\newblock In A.B. Smolders and M.P. van Haarlem, editors, {\em Perspectives on
  Radio Astronomy -- Technologies for Large Antenna Arrays}, 1999.

\bibitem{Bregman:2000}
Jaap~D. Bregman.
\newblock Concept design for a low-frequency array.
\newblock In {\em Proc.SPIE}, volume 4015, pages 4015 -- 4015 -- 14, 2000.

\bibitem{Bregman:2004}
Jaap~D. Bregman.
\newblock {System Optimisation Of Multi-Beam Aperture Synthesis Arrays For
  Survey Performance}.
\newblock {\em Experimental Astronomy}, 17:365--380, June 2004.

\bibitem{Bregman:2012}
Jaap~D. Bregman.
\newblock {\em System Design and Wide-field Imaging Aspects of Synthesis Arrays
  with Phased Array Stations}.
\newblock PhD thesis, Rijksuniversiteit Groningen, 2012.

\bibitem{KPTI-performance}
{KPTI/KAISER Meltdown Initial Performance Regressions}, 2018.

\bibitem{Broekema:18a}
P.~Chris Broekema, J.~Jan~David Mol, Ronald Nijboer, Alexander~S. van
  Amesfoort, Michiel~A. Brenjens, G.~Marcel Loose, and John~W. Romein.
\newblock {Cobalt: a GPU-based correlator and beamformer for LOFAR}.
\newblock {\em Astronomy and Computing}, 23:180--192, April 2018.

\bibitem{Cooley:1965}
James~W Cooley and John~W Tukey.
\newblock An algorithm for the machine calculation of complex {F}ourier series.
\newblock {\em Mathematics of computation}, 19(90):297--301, 1965.

\bibitem{Cornwell:2019}
T.~J. Cornwell, P.~Wortmann, B.~Nikolic, and J.~Farnes.
\newblock {SKA1 SDP Algorithm Reference Library (ARL) Report}.
\newblock Technical report, SDP Consortium, 2019.

\bibitem{CroarkenMary1990Esci}
Mary Croarken.
\newblock {\em Early scientific computing in Britain / Mary Croarken.}
\newblock Oxford science publications. Clarendon Press, 1990.

\bibitem{deVos:2001}
C.~M. de~Vos, K.~Van~der Schaaf, and J.~D. Bregman.
\newblock Cluster computers and grid processing in the first radio-telescope of
  a new generation.
\newblock In {\em Proceedings of the 1st International Symposium on Cluster
  Computing and the Grid}, CCGRID '01, pages 156--, Washington, DC, USA, 2001.
  IEEE Computer Society.

\bibitem{dennard:1974}
Robert~H Dennard, Fritz~H Gaensslen, V~Leo Rideout, Ernest Bassous, and Andre~R
  LeBlanc.
\newblock Design of ion-implanted mosfet's with very small physical dimensions.
\newblock {\em IEEE Journal of Solid-State Circuits}, 9(5):256--268, 1974.

\bibitem{ellram:1995}
Lisa~M Ellram.
\newblock Total cost of ownership: an analysis approach for purchasing.
\newblock {\em International Journal of Physical Distribution \& Logistics
  Management}, 25(8):4--23, 1995.

\bibitem{1966MNRAS.134...87E}
B.~{Elsmore}, S.~{Kenderdine}, and Sir {Ryle}, Martin.
\newblock {The operation of the Cambridge one-mile telescope}.
\newblock {\em Monthly Notices of the Royal Astronomical Society}, 134:87, Jan
  1966.

\bibitem{florio:2016}
Massimo Florio, Stefano Forte, and Emanuela Sirtori.
\newblock {Forecasting the socio-economic impact of the Large Hadron Collider:
  A cost--benefit analysis to 2025 and beyond}.
\newblock {\em Technological Forecasting and Social Change}, 112:38--53, 2016.

\bibitem{graser:2015}
Ferdl Graser and John Taylor.
\newblock {SKA SDP Costing basis of estimate}.
\newblock Technical report, SDP Consortium, 2015.

\bibitem{guzman:2014}
Juan~Carlos Guzman, Gianluca Chiozzi, Alan Bridger, and Jorge Ibsen.
\newblock The cost of developing and maintain the monitoring and control
  software of large ground-based telescopes.
\newblock In {\em Software and Cyberinfrastructure for Astronomy III}, volume
  9152, page 91521P. International Society for Optics and Photonics, 2014.

\bibitem{holties:2012}
H.A. Holties.
\newblock {BP/P Replacement options}.
\newblock Technical report, ASTRON, May 2012.

\bibitem{kemball:2004}
Athol~J Kemball and TJ~Cornwell.
\newblock A simple model of software costs for the square kilometre array.
\newblock {\em Experimental Astronomy}, 17(1-3):317--327, 2004.

\bibitem{Kocher2018spectre}
Paul Kocher, Daniel Genkin, Daniel Gruss, Werner Haas, Mike Hamburg, Moritz
  Lipp, Stefan Mangard, Thomas Prescher, Michael Schwarz, and Yuval Yarom.
\newblock Spectre attacks: Exploiting speculative execution.
\newblock {\em ArXiv e-prints}, January 2018.

\bibitem{Lariviere:18}
Vincent Lariviere and Cassidy~R. Sugimoto.
\newblock The journal impact factor: A brief history, critique, and discussion
  of adverse effects.
\newblock In W.~Glanzel, H.~F. Moed, U.~Schmoch, and M.~Thelwall, editors, {\em
  Springer Handbook of Science and Technology Indicators}. Springer
  International Publishing, Cham, Switzerland, 2018.

\bibitem{Lipp2018meltdown}
Moritz Lipp, Michael Schwarz, Daniel Gruss, Thomas Prescher, Werner Haas,
  Stefan Mangard, Paul Kocher, Daniel Genkin, Yuval Yarom, and Mike Hamburg.
\newblock Meltdown.
\newblock {\em ArXiv e-prints}, January 2018.

\bibitem{mckay:2014}
Derek McKay-Bukowski, Juha Vierinen, Ilkka~I Virtanen, Richard Fallows, Markku
  Postila, Thomas Ulich, Olaf Wucknitz, Michiel Brentjens, Nico Ebbendorf,
  Carl-Fredrik Enell, et~al.
\newblock {KAIRA: The Kilpisj{\"a}rvi atmospheric imaging receiver
  array—System overview and first results}.
\newblock {\em IEEE Transactions on Geoscience and Remote Sensing},
  53(3):1440--1451, 2014.

\bibitem{1990Ahos}
Stephen~G. Nash.
\newblock {\em A history of scientific computing}.
\newblock ACM Press history series. Addison-Wesley, 1990.

\bibitem{currency}
National archives currency converter, 2018.

\bibitem{Needham:1992}
Roger~M. Needham.
\newblock {Later Developments at Cambridge: Titan, CAP, and the Cambridge
  Ring}.
\newblock {\em {IEEE Annals of the History of Computing}}, 14, 1992.

\bibitem{prasad:2016}
Peeyush Prasad, Folkert Huizinga, Eric Kooistra, Daniel van~der Schuur, Andre
  Gunst, John Romein, Mark Kuiack, Gijs Molenaar, Antonia Rowlinson, John~D
  Swinbank, et~al.
\newblock {The AARTFAAC All-Sky Monitor: System Design and Implementation}.
\newblock {\em Journal of Astronomical Instrumentation}, 5(04):1641008, 2016.

\bibitem{Romein:10}
John~W. Romein, P.~Chris Broekema, Jan~David Mol, and Rob~V. van Nieuwpoort.
\newblock {The LOFAR Correlator: Implementation and Performance Analysis}.
\newblock In {\em ACM Symposium on Principles and Practice of Parallel
  Programming (PPoPP'10)}, pages 169--178, Bangalore, India, January 2010.

\bibitem{Romein:06}
John~W. Romein, P.~Chris Broekema, Ellen van Meijeren, {Kjeld van der} Schaaf,
  and Walther~H. Zwart.
\newblock {Astronomical Real-Time Streaming Signal Processing on a Blue Gene/L
  Supercomputer}.
\newblock In {\em ACM Symposium on Parallel Algorithms and Architectures
  (SPAA'06)}, pages 59--66, Cambridge, MA, July 2006.

\bibitem{AboutRS}
About the {Royal Society}, 2018.

\bibitem{ElectRS}
Elections, 2018.

\bibitem{RoyalFellows}
{Fellows of the Royal Society 1660-2007}, 2008.

\bibitem{Ryle:65}
M.~Ryle, B.~Elsmore, and Ann~C. Neville.
\newblock Observations of radio galaxies with the one-mile telescope at
  {C}ambridge.
\newblock {\em Nature}, 207:1024--1027, September 1964.

\bibitem{1962MNRAS.125...39R}
M.~{Ryle} and A.~C. {Neville}.
\newblock {A radio survey of the North Polar region with a 4.5 minute of arc
  pencil-beam system}.
\newblock {\em Monthly Notices of the Royal Astronomical Society}, 125:39, Jan
  1962.

\bibitem{Ryle:1974}
Martin Ryle.
\newblock Radio telescopes of large resolving power.
\newblock In Sven Lundqvist, editor, {\em Nobel Lectures, Physics 1971-1980}.
  World Scientific Publishing Co., Singapore, 1992.

\bibitem{schopper:2016}
Herwig Schopper.
\newblock {Some remarks concerning the cost/benefit analysis applied to LHC at
  CERN}.
\newblock {\em Technological Forecasting and Social Change}, 112:54--64, 2016.

\bibitem{serylak2012}
Maciej Serylak, Aris Karastergiou, Chris Williams, Wesley Armour, Michael
  Giles, LOFAR Pulsar~Working Group, et~al.
\newblock {Observations of transients and pulsars with LOFAR international
  stations and the ARTEMIS backend}.
\newblock {\em Proceedings of the International Astronomical Union},
  8(S291):492--494, 2012.

\bibitem{SKA}
The road to key science observations, 2015.

\bibitem{smith:1988}
James~E Smith.
\newblock Characterizing computer performance with a single number.
\newblock {\em Communications of the ACM}, 31(10):1202--1207, 1988.

\bibitem{Stone}
Richard Stone.
\newblock Input-output and demographic accounting: A tool for educational
  planning.
\newblock {\em Minerva}, IV(3), 1966.

\bibitem{rathenau:2019}
S.Y. Tjong~Tjin Tai, J.~van~den Broek, and J.~Deuten.
\newblock {De impact van grootschalige onderzoeksinfrastructuren -- Een
  meetmethode voor de return on investment van internationale
  onderzoeksfaciliteiten}, 2019.
\newblock In Dutch.

\bibitem{taylor:2019}
J.~Taylor.
\newblock {SKA1 SDP Performance Prototype Platform (P3-Alaska) Prototyping
  Report}.
\newblock Technical report, SDP Consortium, 2019.

\bibitem{Schaaf:2003}
K.~van~der Schaaf, J.~D. Bregman, and C.~M. de~Vos.
\newblock Hybrid cluster computing hardware and software in the {LOFAR} radio
  telescope.
\newblock In {\em Proceedings of the International Conference on Parallel and
  Distributed Processing Techniques and Applications, {PDPTA}}, volume~2, pages
  695--701, Las Vegas, Nevada, USA, June 2003.

\bibitem{Schaaf:2004}
Kjeld Van Der~Schaaf and Ruud Overeem.
\newblock {COTS} correlator platform.
\newblock {\em Experimental Astronomy}, 17(1):287--297, June 2004.

\bibitem{haarlem:2013}
M.~P. van Haarlem, M.~W. Wise, A.~W. Gunst, et~al.
\newblock {LOFAR: The LOw-Frequency ARray}.
\newblock {\em {Astronomy and Astrophysics}}, 556, August 2013.

\bibitem{Nieuwpoort:10}
Rob~V. van Nieuwpoort and John~W. Romein.
\newblock {Building Correlators with Many-Core Hardware}.
\newblock {\em IEEE Signal Processing Magazine (special issue on "Signal
  Processing on Platforms with Multiple Cores: Part 2 -- Design and
  Applications")}, 27(2):108--117, March 2010.

\bibitem{incose:2015}
David~D Walden, Garry~J Roedler, Kevin Forsberg, R~Douglas Hamelin, and
  Thomas~M Shortell.
\newblock {\em Systems engineering handbook: A guide for system life cycle
  processes and activities}.
\newblock John Wiley \& Sons, 2015.

\bibitem{WilkesM.V.MauriceVincent1985Moac}
M.~V. (Maurice~Vincent) Wilkes.
\newblock {\em Memoirs of a computer pioneer / by Maurice V. Wilkes.}
\newblock The MIT Press series in the history of computing. MIT Press,
  Cambridge, Mass. ; London, 1985.

\bibitem{wouters:2005}
Marc Wouters, James~C. Anderson, and Finn Wynstra.
\newblock The adoption of total cost of ownership for sourcing decisions––a
  structural equations analysis.
\newblock {\em Accounting, Organizations and Society}, 30(2):167 -- 191, 2005.

\bibitem{zarka:2015}
P.~Zarka, Mohammed Nabil~El Korso, Remy Boyer, Pascal Larzabal, et~al.
\newblock {NenUFAR: Instrument description and science case}.
\newblock {\em 2015 International Conference on Antenna Theory and Techniques
  (ICATT)}, pages 1--6, April 2015.

\end{thebibliography}
\end{document}